\documentstyle[12pt]{article}
 
  \def\n{\nonumber}
  
  \def\th{\theta}
  \def\a{\alpha}
  \def\b{\beta}
  \def\be{\begin{equation}}
  \def\ee{\end{equation}}
  \def\bq{\begin{eqnarray}}
  \def\eq{\end{eqnarray}}
  \def\p{\partial}
  \def\({\left(}
  \def\){\right)}
  \def\lb{\left[}
  \def\rb{\right]}
  \title{\bf\huge Gravoelectric-dual of the Kerr solution}
  \author{Naresh Dadhich\thanks{E-mail : nkd@iucaa.ernet.in} \\
{\sl Inter-University Center for Astronomy \& Astrophysics,}\\
{\sl Post Bag 4, Ganeshkhind, Pune - 411 007, India.} \\
 L.K. Patel \\
 {\sl Department of Mathematics,  }\\
{\sl Gujarat University, Ahmedabad 380 009, Gujarat, India.}}

\begin{document}
\maketitle
\begin{abstract}

  By decomposing the Riemann curvature into electric and magnetic 
parts, we define the gravoelectric duality transformation by interchange 
of active and passive electric parts which amounts to interchange of the
Ricci and Einstein tensors. It turns out that the vacuum equation is 
duality-invariant. We obtain solutions dual to the 
Kerr solution by writing an effective vacuum equation in such a way that 
it still admits the Kerr solution but is not duality invariant. The dual 
equation is then solved to obtain the dual-Kerr solution which 
can be interpreted as the Kerr black hole sitting in a string dust universe.

\end{abstract}
  
 
% \begin{keyword}
  {\it Duality transformation, Kerr solution, global monopole, string dust.} 
%  \end{keyword}
\newpage  
  %\end{frontmatter}
%\baselineskip=24pt
  
\noindent{\bf I. Introduction}
 
  In analogy with the Maxwell electromagnetic field, it is also 
possible
 to resolve gravitational field;i.e. the Riemann curvature into electric and
 magnetic parts relative to a timelike observer. In general relativity (GR), 
there are two kinds of gravitational charges; one non-gravitational energy 
distribution and the other gravitational field energy. Thus electric part 
would have further decomposition into active and passive parts corresponding 
to these two kinds [1-3]. Electromagnetic parts would be given by 
second rank tensors orthogonal to the resolving unit timelike vector. Electric 
parts are symmetric and account for 12 (6 each for active and passive) 
while magnetic part is trace free and accounts for the remaining 8 components 
of the Riemann curvature. The symmetric part of the magnetic part is 
equal to the Weyl magnetic part and antisymmetric part represents energy 
flux.

 We consider a transformation which is interchange of active and 
passive electric parts [1] and we term it as electrogravity duality. It 
turns out that the vacuum equation is symmetric in active and passive 
parts and hence is duality invariant. The vacuum solutions would thus remain 
invariant modulo sign of constants of integration. It does in fact 
happen that the Riemann curvature for vacuum solutions changes sign under 
the duality transformation. That means $GM \longrightarrow 
-GM$ in the Schwarzschild solution. Thus vacuum solutions are self-dual 
modulo change of sign of $G$. This can however be understood as follows. 
The active part is anchored on the 
non-gravitational energy distribution while the passive part on the 
gravitational field energy [4]. For an attractive field the former is 
positive while the latter is negative and hence interchange of active and 
passive parts must naturally require $G\longrightarrow -G$. 

 Now the question arises, can the symmetry of the equation be broken 
to get distinct dual solutions? It turns out that in obtaining 
the well-known and physically interesting black hole solutions, there 
always remains one equation unused which is implied by the others [4]. If we 
tamper this equation, vacuum solutions would remain undisturbed and the 
symmetry between and active and passive parts would be broken. This is 
precisely what happens [1,3,4], and it is then possible to obtain distinct 
dual solutions. That is, it is possible to write an effective vacuum 
equation in such 
a way that it gives the same vacuum solutions but it is not 
duality invariant. Then the dual equation would yield distinct solutions.. 
Following this method, solutions 
dual to Schwarzschild, Reissner-Nordstr$\ddot o$m, and NUT solutions have 
been obtained [1,3,5,6]. A dual solution is obviously non-vacuum and it is 
also asymptotically non-flat. It gives rise to energy momentum distribution 
which agrees with string dust distribution or that of a global monopole at 
large distance from the core. Thus duality transformation imbibes string dust 
or global monopole charge automatically, and dual spacetime describes the 
original source sitting in a string dust universe [7-8] or having a global 
monopole charge [9].

 Note that the duality transformation interchanges  active and 
passive electric parts which are respectively (double) projections of the 
Riemann and its double (left and right) dual onto a timelike observer. 
Since contraction of the Riemann is the Ricci while that of its double 
dual is the Einstein tensor, the duality transformation would hence imply 
interchange between the Ricci and Einstein tensors. The electrogravity 
duality implies the Ricci-Einstein duality. Thus 
the dual equation would result when the Ricci components are replaced by 
the Einstein components in the effective vacuum equation.

 In this paper by application of the duality transformation we wish to 
obtain spacetime dual to the Kerr rotating black 
hole. As expected it would be  quite involved and different from the 
other cases. Firstly, manipulation of equations is very complicated and 
secondly the dual equation, unlike the other cases, admits more than one 
solutions. That is dual solution is not unique. Dual spacetimes could 
represent the Kerr black hole in a string dust universe, as the stresses 
generated by the duality conform with that of string dust. They admit a 
horizon but it would not be a $r = const.$ but instead be 
$r=f(\theta)$ surface. When string dust density is switched off 
they reduce to the Kerr black hole.

 In Sec.II we decompose the Riemann curvature in electromagnetic parts and 
then write the vacuum equation and consider the duality transformation. 
This is followed in Sec.III by solving the dual equation for the axially 
symmetric metric to obtain dual solutions. In Sec.IV we discuss the 
string dust interpretation and conclude in Sec.V with discussion on the 
general aspects of the duality transformation.

\noindent{\bf II. Electromagnetic decomposition and duality}
 
  We first resolve the Riemann curvature relative to a timelike unit vector
 into electromagnetic parts as follows [1],
 
  \be
      E_{ac} = R_{abcd} u^b u^d,  \tilde E_{ac} = *R*_{abcd} u^b u^d
      \ee
      
      \be
      H_{ac} = *R_{abcd} u^b u^d = H_{(ac)} - H_{[ac]}
      \ee
      
      \n where
      
      \be
      H_{(ac)} = *C_{abcd} u^b u^d,~ 
      H_{[ac]} = \frac{1}{2} \eta_{abce} R^e_d u^b u^d.
      \ee

  \noindent Here $C_{abcd}$ is the Weyl conformal curvature, $\eta_{abcd}$ is the
 4-dimensional volume element. Note that $E_{ab}$ and $\tilde E_{ab}$ are
 symmetric, $H_{ab}$ is trace-free and they are all orthogonal to the
 resolving unit timelike vector $u^a$. In terms of them, the Ricci
 curvature reads as 
 
  \bq
  R_{ab} &=& E_{ab} + \tilde E_{ab} + (E + \tilde E)u_au_b - \tilde E
 g_{ab} + H^{mn}(\eta_{acmn}u_b + \eta_{bcmn}u_a) u^c. 
\label{4} \eq
 
  \noindent Then the vacuum equation for any timelike unit resolving vector would 
imply
 
  \bq
  E ~or~ \tilde E = 0 ,  H_{[ab]} = 0  , E_{ab} + \tilde E_{ab} = 0
 \label{5} \eq
 
 \noindent which is symmetric in $E_{ab}$ and $\tilde E_{ab}$.

  \noindent We define the electrogravity duality transformation by 
 
  \bq
  E_{ab} \leftrightarrow \tilde E_{ab},  H_{[ab]} &=& H_{[ab]}. \label{6} 
\eq
 
 \noindent Obviously the vacuum equation is symmetric in $E_{ab}$ and $\tilde 
E_{ab}$ and hence would remain invariant under the duality 
transformation. It would give rise to the same vacuum solution. It 
turns out that the Weyl curvature changes sign under 
the duality [2], which means the constants of integration representing 
physical parameters like mass, angular momentum and NUT charge in the 
vacuum solutions must change sign. It is equivalent to $G \longrightarrow
- G$. It can in fact be verified for the 
Kerr solution  by looking at the Riemann components as given in 
[10]. The vacuum solutions would modulo sign of $G$ be self dual.

 \noindent It turns out that while obtaining the Schwarzschild solution, there 
remains one equation unused, which is implied by the others. In 
particular, $H_{[ab]} = 0$, $\tilde E = 0$ and $E_{22} + \tilde E_{22} = 0$ 
determine the 
solution completely leaving $E_{11} + \tilde E_{11} = 0$ free. In terms 
of the Ricci components, this is equivalent to $R_{01} = 0 = R_{22}, 
R^0_0 = R^1_1$ with $R_{00} = 0$ being free,  which is implied by the 
others. That is the set $R_{01} = 0 = R_{22}, R^0_0 = R^1_1$ suffices to 
give the unique Schwarzschild solution. We take this as the effective vacuum 
equation which also yields the Schwarzschild solution uniquely. The set dual 
to the effective vacuum equation would be obtained by changing Ricci to 
Einstein,viz $G_{01} = 0 = G_{22}, G^0_0 = G^1_1$. This dual set then 
admits the unique solution which can be interpreted as the 
Schwarzschild black hole with a global monopole charge [9] or sitting in 
a string dust universe [7-8]. By this method of the duality transformation, 
the solutions dual to the Reissner-Nordstr$\ddot o$m and the NUT solutions 
have also been obtained [1,3,5,6].

 \noindent In the next section we wish to apply this method to axially 
symmetric 
spacetime to obtain solution dual to the Kerr solution. It turns out that 
in this case the solution is not unique and we find two distinct 
solutions.

\noindent{\bf III. Dual solutions}
  
  We consider an axially symmetric line element in the form [10],
 
 \bq 
  ds^2 &=& 2 (du + g \sin\a d\b) dt - M^2 (d\a^2 + 
  \sin^2\a d\b^2)  \cr
  &&\mbox{} - 2 L (du + g \sin\a d\b)^2. \label{7} 
\eq

  Here $M$ and $L$ are functions of $u$, $\a$ and $t$ and $g$ is a 
  function of
  $\a$ only.  We use $u$, $\a$, $t$ and $\b$ as coordinates. \\ 
  Introducing the tetrads

  \bq
  \th^1 &=& d u + g \sin \a d \b, \qquad \quad \th^2 = M d \a,\n \cr
  \th^3 &=& M \sin \a d \b, \qquad \quad \th^4 = d x - L \th^1 
   \n \eq 

  we can express the metric (7) in the form
  \be 
  ds^2 = 2 \th^1 \th^4 - (\th^2)^2 - (\th^3)^2 = g_{(ab)} \th^a 
  \th^b.
  \label{8} \ee
  The components $R_{ab}$ of the Ricci tensor for the metric 
  (\ref{7}) were
  obtained by Vaidya {\it et al} [11].  We reproduce them below for 
  ready reference, and they are given in the basis frame as follows:
  \bq
  R_{23} &=& 0 \n \\
  R_{44} &=& \frac{2}{M}\left[M_{xx} - \frac{f^2}{M^3}\right] \n \\
  R_{24} &=& \frac{g}{M} \lb \(\frac{M_x}{M}\)_y - 
  \(\frac{f}{M^2}\)_u\rb
  \n \\
  R_{34} &=& - \frac{g}{M} \lb \(\frac{M_x}{M}\)_u - 
  \(\frac{f}{M^2}\)_y\rb
  \n \\
  R_{14} &=& \frac{2}{M} \lb M_{xu} + \(LM_x\)_x + \(\frac{L 
  f^2}{M^3}\)\rb
  + L_{xx} \label{9} \\
  R_{12} &=& L R_{24} + \frac{g}{M} \lb\(L_x + \frac{M_u}{M}\)_y + 
  \(
  \frac{2fL}{M^2}\)_u\rb \n \\
  R_{13} &=& L R_{34} + \frac{g}{M} \lb-\(L_x + \frac{M_u}{M}\)_u + 
  \(
  \frac{2fL}{M^2}\)_y\rb \n \\
  R_{22} &=& R_{33} = \frac{1}{M^2} \lb g^2 \(\frac{M_u}{M}\)_u + 
  g^2
  \(\frac{M_y}{M}\)_y - 1 + 2 f \(\frac{M_y}{M}\) \right. \n \\
  &&\left.\mbox{}+ 4 \(\frac{f^2L}{M^2}\) -
  \(M^2\)_{ux} - \( L \(M^2\)_x\)_x\rb \n \\
  R_{11} &=& L^2 R_{44} + \frac{1}{M^2}\lb g^2 (L_{uu} + L_{yy}) + 
  2 f
  L_y + 2 L_u M M_x + 4 L M M_{xu} \right. \n \\
  &&\left.\mbox{} - 2 L_x M M_u + 2 M M_{uu}\rb. \n
  \eq
  In the above equations, the variable $y$ replaces $\a$, the defining 
  relation
  being

\be 
g d\a = dy. \label{10} 
\ee

  Here and in what follows a suffix denotes partial differentiation, 
  eg.,
  $g_{\a} = \p g/\p\a$, $L_y = \p L/\p y$, $M_{xu} = \p^2M/\p x \p u$ 
  etc and $x = t$.  The
  symbol $2f$ stands for the expression $g_{\a} + g \cot\a$.

 \noindent In the case of spherical symmetry it was $R_{00} = 0$ was free while 
in this case for the metric (7) it is 
$R_{14} = 0$ is free. That is it is implied by the others and is not used in 
obtaining the vacuum solution. We shall thus consider the effective vacuum 
equation as
\bq
R_{ab} = 0,  ~except R_{14}. \label{11}
\eq
\noindent It can be verified from (9) that solution of the rest of the 
equation automatically implies $R_{14} = 0$ giving the vacuum Kerr solution.
   
 \noindent The dual equation would be obtained by letting $R_{ab}\rightarrow 
G_{ab}$ and hence it would be
\bq
G_{ab} = 0,  ~except G_{14},\label{12} 
\eq
\n which would imply
\bq
R_{ab} = 0,  ~except R_{22} = R_{33}. \label{13}
\eq
 
 \noindent To find the dual solution we have to solve the set (13) which would 
read as 
  \bq
  R_{44} &=& 0, \qquad R_{24} = 0, \qquad R_{34} = 0 \label{14} 
  \\
  R_{14} &=& 0 \label{15} \\
  R_{12} &=& 0, \qquad R_{13} = 0 \label{16} \\
  R_{22} &=& -\rho\label{17} \\
  R_{11} &=& 0. \label{18} \eq
 
  We proceed as follows.\\
 
 Eqns. (14-18) yield the general solutions
 
  \be M^2 = \frac{f}{Y} (X^2 + Y^2) \label{19} \ee
  and
  \be 2 L = - \frac{Y_u}{Y} X + 2 G + \frac{2 A X + 2 B Y}{X^2 + Y^2}, 
  \label{20}
  \ee
  where $X$ is a function of $t$, $u$ and $y$ and $Y$, $A$, $B$ and $G$ 
  are functions of $u$ and $y$ satisfying the relations
  \bq
  X_t &=& -1, \qquad X_y = Y_u, \qquad X_u = - Y_y \label{21} \\
  B &=& - 2 Y G - Y Y_y \label{22}
  \eq
  and
  \be B_u = A_y, \qquad B_y = - A_u. \label{23} \ee
  Eqn. (19) determines 

\be 
\rho = - \frac{1}{X^2+Y^2}\lb 2 G + \frac{Y}{f}\(\frac{1}{2} g^2
\nabla^2 \log\(\frac{Y}{f}\) - f_y + 1 + 3f \frac{Y_y}{Y}\)\rb, 
\label{24} 
\ee

  where $\nabla^2 \equiv \p^2/\p u^2 + \p^2/\p y^2$. \\  
 
  Now we assume that $Y$ is a function of $y$ only. This leads from
 eqn.(21) to
  \be X =   a u - t, \qquad Y = - a y + b, \label{25} \ee
  where $a$ and $b$ are constants of integration, no additional 
  constant is
  added in $X$ because such a constant can always be incorporated in 
  the
  $t$--coordinate.
  
  Eqns. (22), (23) and (25) will then lead to
  \be Y \nabla^2 G - 2 a G_y = 0 \label{26} \ee
  of which we take the particular solution

  \be 2 G = \mbox{constant } = c. \label{27} \ee
  Eqns. (22) and (25) then lead  to
  \be B = (a -c) Y, \qquad A = a(a-c) u + m, \label{28}
  \ee

  \noindent where $m$ is again a constant of integration.
  
  We now introduce the variable $\th$ and a function $h(\th)$ as 
  follows:

  \be 
  \(\frac{f}{Y}\)^{1/2}d\a = d \th, \qquad 
  \(\frac{f}{Y}\)^{1/2}
  \sin\a = h(\th) \label{29}. 
  \ee

  Using (10), (25) and (29) we can show that
  
\be 
\frac{Y}{f} \left\{ g^2 \nabla^2 \log\(\frac{Y}{f}\) - f_y 
+ 1 + 3 f
\frac{Y_y}{Y}\right\} = - 2 a - \frac{h_{\th\th}}{h}. \label{30} 
\ee
  
  The Kerr solution would follow if we take $h_{\th\th}/h = -1$, which would 
 imply $f = Y$ and give 
  \be
  2 L = c + \frac{2X[ a (a-c)u + m] + 2 (a-c)Y^2}{X^2 + Y^2}. 
  \label{31} 
 \ee
 
 So far we have not used eqn.(18), which would now imply $a = c$, 
 and finally we shall have 
 
 \be 
  \rho = \frac{1 - a}{X^2 + Y^2}, \label{32}
 \ee
 
  \noindent where $X$ and $Y$ are given by (25).
  
  Using eqns. (10), (25) and (29) we can 
  write
  \be 2 h Y = (g\sin\a)_{\th}, \qquad h Y_{\th} = - a g \sin\a, 
  \label{33} \ee
  where $h = sin\th$.  These relations together then give 
  the following differential equation for $Y$:
  \be Y_{\th\th} + \frac{h_\th}{h} Y_{\th} + 2 a Y = 0. \label{34} \ee
 
  This integrates to give $Y = k cos\th$ for $a = 1$, when $\rho$ 
would vanish giving the Kerr solution. This is the Legendre equation 
which could give non-vacuum dual solution only when the integer $a \neq 
1$ but the solution would not include the Kerr solution as a 
particular case. This would however be a dual solution for different 
integer values of $a$.\\

 If we wish to have a dual solution that includes 
the Kerr solution as a particular case, we will have to give up the 
relation $f = Y$. In that case we can have the following two different 
solutions:

\bq
Y = k cos\th, ~h = sin^{2a - 1}\th, ~g sin\a = \frac{k}{a} 
sin^{2a}\th, \n \\
\rho = \frac {a - 1} {X^2 +
Y^2}  [1 - 2(2a - 1) cot^2\th]. \label{35} 
\eq

\n and

\bq 
Y = k cos^a\th, ~h = sin\th cos^{a-1}\th, ~g sin\a = k sin^2 \th, \n \\
\rho = \frac {a - 1}{X^2 + Y^2} [2 - (a - 2) tan^2\th ]. \label{36} 
\eq

\noindent These are two distinct solutions. Thus dual solution to the Kerr 
solution is not unique. Of course each has stresses corresponding to  a 
string dust distribution of density $\rho$, which will diverge at 
$\th = 0$ for the 
former and at $\th = \pi/2$ for the latter. Both however reduce to the Kerr 
solution when $a = 1$. The metric would read as
  
\bq 
ds^2 &=& 2 (du + g\sin\a d\b)dt - (R^2 + Y^2)(d\a^2 + 
H^2(\th)d\b^2) \n \\ 
&&\mbox{}- \lb a + \frac{2 mR}{R^2 + Y^2} \rb (du + g\sin\a
d\b)^2,
\label{37} \eq

\noindent where we have defined $X = R = (a - 1)t - ar$ as a new radial 
coordinate. For the two dual solutions $Y, h, g\sin\a$ as given above in 
(35-36). The solutions go over to the Kerr solution 
when $a = 1$ with $m$ and $k$ as mass and specific angular momentum.

 The dual solutions do however admit horizon defined by the equation,  
 \be
 a h^2 (R^2 + Y^2) - 2m R h^2 + (a g sin\a)^2 = 0 \label{38}
 \ee
 
\noindent which would for the solution (\ref{35}) give the horizon as

\be
 R_+ = \frac{m}{a} (1 + \sqrt{1 - (k^2/m^2)(a^2 cos^2\th + sin^2\th)}. 
\label{39} 
\ee

\noindent Thus horizon has unlike the Kerr black hole $\th$ dependence. This 
is so for the other solution as well. In other dual solutions [1,3,5], 
the basic character of the field remained unaltered while here it is not 
so as indicated by the $\th$ dependence of the horizon. In the next section 
we shall show that these solutions could be interpreted as rotating black hole
sitting in a string dust universe with the string dust density $\rho$ given 
by (35-36). It is true that the solutions obtained are rather complicated 
and unfortunately cannot be transformed to the Boyer-Lindquist form to 
gain more physical insight. Since dual solution is not 
unique, it may be possible to find a simpler and physically more 
transparent solution. The search is on. \\

\noindent{\bf IV. String dust interpretation}

\noindent In terms of the electromagnetic parts, the effective vacuum equation 
(11) will take the form
  \bq
 \tilde E = 0  , H_{[ab]} = 0 , E_{ab} + \tilde E_{ab} = - (E + \tilde 
E)w_aw_b \label{40} 
\eq
\noindent which is no longer symmetric in active and passive parts. Here 
$w_a$ is a unit spacelike vector 
orthogonal to $u_a$ and in the direction of acceleration vector. It would 
admit the same vacuum solution but it is now not 
invariant under the duality transformation (6).

 The dual equation (12) would read as
 
  \bq
  E = 0 , H_{[ab]} = 0 , ~E_{ab} + \tilde E_{ab} = - (E + \tilde E)w_aw_b  
\label{41}
 \eq
  \noindent which is equivalent to the set (14-18). This is what we have 
solved for dual solution and it gives rise to the only surviving stress 
component
 \bq
T_{14} = \rho \label{42}
\eq 
\noindent where $\rho$ is given by (35) or (36).

 On the other hand a string dust distribution is characterized by [7,8],
  \bq
   T_{ab} &=& \rho (u_au_b - w_aw_b). \label{43} \eq

 For the metric (7) we choose  the timelike and spacelike vectors 
as follows:
\be 
u_{(a)} = (1,0,0,\frac{1}{2}) \qquad w_{(a)} = (1,0,0,-\frac{1}{2}).
\label{44} \ee
  
 Then eqn. (43) would imply the distribution (42). Thus 
our dual solutions could represent a Kerr black hole sitting in a string 
dust universe, and the string density which is equal to radial tension is 
given by (35) and (36) for the two solutions. When string density is 
switched off, the Kerr solution follows.

\noindent{\bf V. Discussion}

 The very first time one encounters the duality transformation is in 
electrodynamics where it prescribes a relation between electric and 
magnetic fields under which the vacuum Maxwell equation remains 
invariant. A similar kind of relationship between active and passive 
electric parts and magnetic part of the gravitational field in fact leads to 
the Einstein vacuum equation [2]. This is because electromagnteic parts 
for gravity stem from the Riemann curvature and hence contain second 
derivative of 
the metric and consequently dynamics of the field. In the Maxwell theory, 
they contain first derivative of the gauge potential, and to get to the 
equation of motion  they need to be differentiated once. There is thus 
a basic difference between electromagnetic parts of the gravitational field 
and that of the Maxwell field. This basic difference should always be 
kept in mind.

 A dual is only defined for an antisymmetric tensor by the Hodge 
dual. The duality transformation represents in general a rotation. 
However we are here considering a relation between active and passive 
electric parts which are symmetric second rank 3-tensors. Thus our 
electrogarvity duality transformation does not represent a duality 
rotation [12] and is therefore different in character from the duality 
transformations considered in other context in GR [13] as well as in other 
theories including the string theory [14]. The duality is at the center 
stage of the current field theory research and has played 
very important role in connecting different theories and situations.

 At any rate, our duality though different from the usual duality, it is 
however a relation between active and passive electric parts and marks 
a symmetry of the vacuum equation. It is thus a valid statement. The 
remarkable feature of this is in finding new dual solutions which imbibe 
global monopole  or string dust automatically. Its connection with 
production of topological defects is rather intriguing and interesting, 
and this feature permeates in lower and higher dimenstions as well as in 
scalar tensor theories [15-17]. 
In the case of the Schwarzschild field, the duality simply restores the 
gauge freedom in choosing zero of the potential one had in the Newtonian 
theory. The vacuum equation does not permit this freedom becuase  the 
spacetime is asymptotically flat and the potential could only vanish at 
infinity. This means that topological defects (dual solutions) thus donot 
disturb the basic 
character of the field at the Newtonian level. However we donot yet fully 
understand the physical meaning and import of the electrogravity duality. 

 Physical features of the black hole with global monopole charge [9]  
have been considered by several authors [18-21]. It has been argued that 
since the global monopole solution (and so are the dual solutions in 
general), is not asymptotically flat, hence its asymptotic regions would be 
curved. Note that positivity of ADM mass is proved only for isolated system 
that generates asymptotic Minkowski geometry. Objects with negative mass may 
therefore generate non-flat asymptotic regions [20]. This suggests an 
association of duality with negative mass. Recall that we have discussed 
in the Introduction that duality would imply gravitational constant 
turning negative which means gravitational mass turning negative. This 
is because interchange of active and passive parts would imply 
interchange of positive  non-gravitational matter energy and negative 
gravitational field energy. This seems consistent with the strange and 
unusual thermodynamical behaviour of black holes with gauge cosmic strings and 
global monopoles [20-21]. There appears to be a deep connection between the 
electrogravity duality and the topological defects. It calls for 
a comprehensive study to probe it further and deeper.

\noindent{\bf Acknowledgement :} It is a pleasure to thank the anonymous referee 
for many helpful comments that have improved the presentation. LKP thanks 
IUCAA for hospitality.

\end{document}